\documentclass[journal=jpclcd,manuscript=letter]{achemso}
\usepackage{booktabs}
\usepackage{commath}

\makeatletter

\author{Jiachen Li}
\affiliation{Department of Chemistry, Yale University, New Haven, CT, USA 06520}
\email{jiachen.li@yale.edu}
\author{Yu Jin}
\affiliation{Department of Chemistry, University of Chicago, Chicago, Illinois 60637, United States}
\author{Jincheng Yu}
\affiliation{Department of Chemistry, Duke University, Durham, NC 27708, USA}
\author{Weitao Yang}
\affiliation{Department of Chemistry, Duke University, Durham, NC 27708, USA}
\email{weitao.yang@duke.edu}
\author{Tianyu Zhu}
\affiliation{Department of Chemistry, Yale University, New Haven, CT, USA 06520}
\email{tianyu.zhu@yale.edu}

\title{Accurate Excitation Energies of Point Defects from Fast Particle-Particle Random Phase Approximation Calculations}

\makeatother

\begin{document}

\begin{tocentry}
\includegraphics[width=1\textwidth]{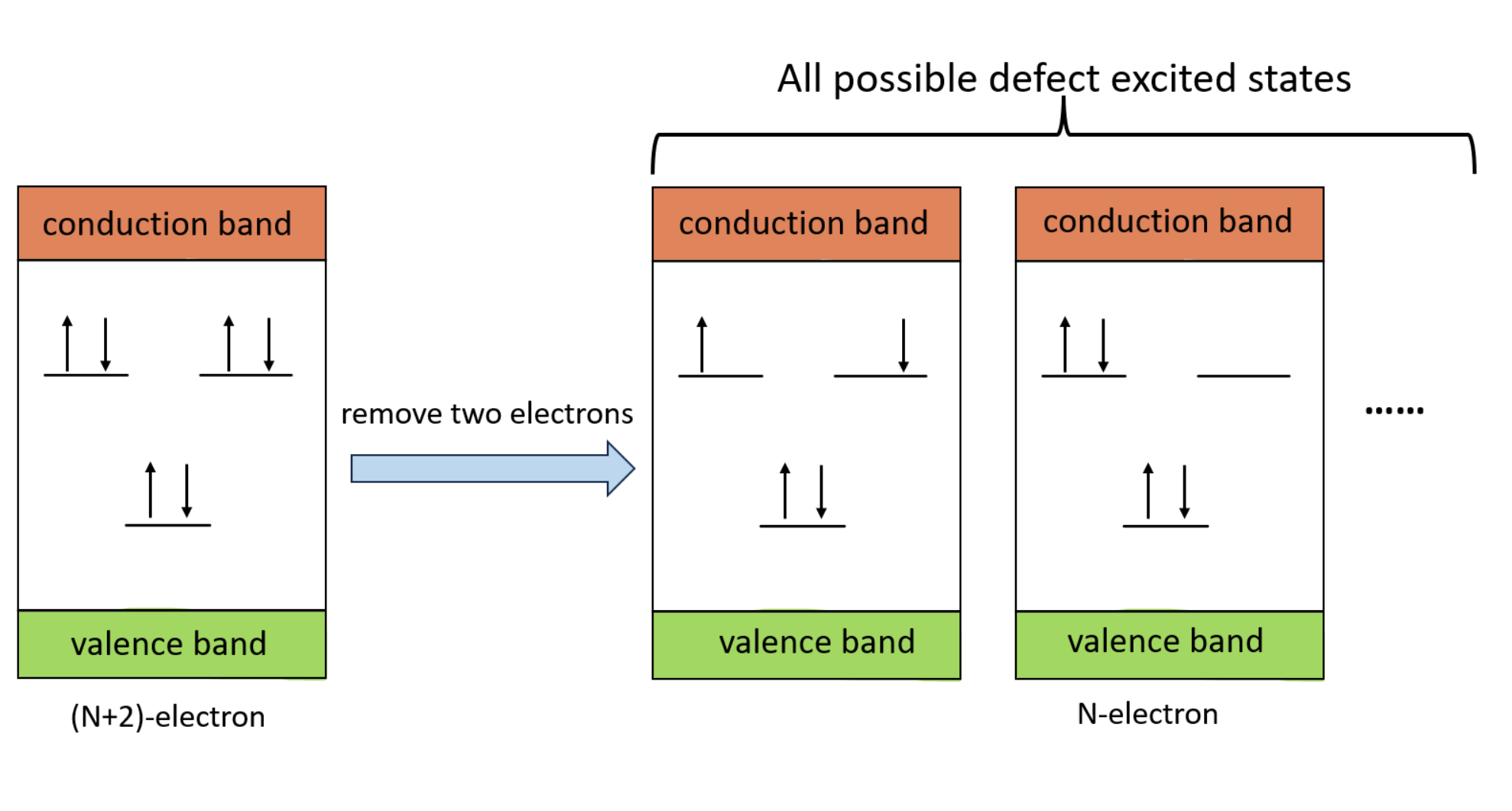}
\end{tocentry}

\begin{abstract}
We present an efficient particle-particle random phase approximation (ppRPA) approach that predicts accurate excitation energies of point defects, 
including the nitrogen-vacancy (NV$^-$) and the silicon-vacancy (SiV$^0$) centers in diamond and the divacancy center (VV$^0$) in 4H silicon carbide, 
with errors within 0.2 eV compared with experimental values. 
Starting from the ($N+2$)-electron ground state calculated with the density functional theory (DFT), 
the ppRPA excitation energies of the $N$-electron system are calculated as the differences between the two-electron removal energies of the ($N+2$)-electron system.  
We demonstrate that the ppRPA excitation energies converge rapidly with a few hundred of canonical active-space orbitals. 
We also show that active-space ppRPA has weak DFT starting-point dependence and is significantly cheaper than the corresponding ground-state DFT calculation. 
This work establishes ppRPA as an accurate and low-cost tool for investigating excited-state properties of point defects and opens up new opportunities for applications of ppRPA to periodic bulk materials. 
\end{abstract}

Optically addressable point defects in semiconductors and insulators have gained 
increasing attention due to their unique optical and magnetic properties and promises for 
realizing quantum technologies~\cite{dreyerFirstPrinciplesCalculationsPoint2018,
wolfowiczQuantumGuidelinesSolidstate2021,
awschalomQuantumTechnologiesOptically2018}. 
In many point defect systems, 
the energy levels of the defect are within the fundamental band gap of the host material, 
which leads to long spin coherence time even at room temperature. 
The spin states of point defects can also be initialized, manipulated, 
and read out through optical excitations. 
As a result, 
point defects have the potential to function as quantum bits (qubits) for quantum computation 
and single-photon emitters for quantum communication~\cite{weberQuantumComputingDefects2010,
jelezkoObservationCoherentOscillation2004,
andersonFivesecondCoherenceSingle2022,
christleIsolatedSpinQubits2017}. 
However, predicting optical properties of point defects from first principles remains 
challenging~\cite{ivadyFirstPrinciplesCalculation2018}. 
Due to the fact that defect excited states often have multiconfigurational character, 
a balanced treatment of static and dynamical electron correlation
is required for accurately describing these states. 
Moreover, large supercells with hundreds of atoms are typically required to 
avoid interactions between periodic images and simulate in the dilute limit, 
which is not feasible for correlated quantum chemistry methods. 

Due to high computational efficiency, 
excited-state extensions of the density functional theory (DFT)~\cite{hohenbergInhomogeneousElectronGas1964,
parrDensityFunctionalTheoryAtoms1989,
kohnSelfConsistentEquationsIncluding1965}, 
including spin-conserving and spin-flip time-dependent DFT (TDDFT) as well as $\Delta$SCF, 
have been widely used to describe optical properties of point defects such as 
vertical excitation energies, zero-phonon lines, optical spectra, and excited-state geometries
\cite{galiInitioSupercellCalculations2008,
maExcitedStatesNegatively2010,
delaneySpinPolarizationMechanismsNitrogenVacancy2010,
jinExcitedStateProperties2023,
jinPhotoluminescenceSpectraPoint2021,
jinVibrationallyResolvedOptical2022}.
However, 
it is challenging for the single-determinant-based DFT formalism to describe 
the strongly correlated defect states with multiconfigurational character. 
Moreover, TDDFT has an undesired significant starting-point dependence on the exchange-correlation functional. 
Periodic many-body treatments of defect excited states have been explored using 
the Bethe-Salpeter equation (BSE) combined with $GW$~\cite{gaoQuasiparticleEnergiesOptical2022,
choiMechanismOpticalInitialization2012}, 
the equation-of-motion coupled-cluster theory~\cite{galloPeriodicEquationofmotionCoupledcluster2021},
as well as the quantum Monte Carlo approach\cite{simulaCalculationEnergiesMultideterminant2023}.
However, the high computational costs have limited further applications of these methods.
To describe the correlated defect systems with affordable cost, 
many flavors of quantum embedding approaches have been developed.
In quantum embedding theories,
a chosen active space representing the manifold of defect states is treated by 
accurate but computationally demanding high-level theories, 
while the rest of the system is treated by cheaper low-level theories~\cite{sunQuantumEmbeddingTheories2016,
cuiEfficientImplementationInitio2020,
zhuEfficientFormulationInitio2020,
zhuInitioFullCell2021}.
It has been shown that the quantum defect embedding theory (QDET)~\cite{maQuantumSimulationsMaterials2020,
maQuantumEmbeddingTheory2021,shengGreenFunctionFormulation2022}, 
the density matrix embedding theory (DMET)~\cite{haldarLocalExcitationsCharged2023,
mitraExcitedStatesCrystalline2021,vermaOpticalPropertiesNeutral2023},
the constrained random phase approximation (CRPA) 
combined with exact diagonalization (ED)~\cite{bockstedteInitioDescriptionHighly2018,
muechlerQuantumEmbeddingMethods2022}
and the regional embedding theory~\cite{lauOpticalPropertiesDefects2023} have achieved mixed successes for simulating point defect systems. Recently, a dynamical downfolding approach has been developed to treat localized correlated electronic states in the otherwise weakly correlated host medium\cite{romanovaDynamicalDownfoldingLocalized2023}.

The particle-particle random phase approximation (ppRPA) formalism, 
originally developed for describing nuclear many-body interactions\cite{
ringNuclearManybodyProblem1980,
ripkaQuantumTheoryFinite1986},
has been successfully applied to describe ground-state electron correlation as well as excited-state energies and oscillator strengths 
in quantum chemistry\cite{vanaggelenExchangecorrelationEnergyPairing2013,vanaggelenExchangecorrelationEnergyPairing2014,yangDoubleRydbergCharge2013}.
ppRPA can be derived from different approaches, 
including 
the equation of motion\cite{ringNuclearManybodyProblem1980,
roweEquationsofMotionMethodExtended1968}, 
the adiabatic connection\cite{vanaggelenExchangecorrelationEnergyPairing2013,
vanaggelenExchangecorrelationEnergyPairing2014} 
and TDDFT with the pairing field\cite{pengLinearresponseTimedependentDensityfunctional2014}.
As the counterpart of the commonly used particle-hole random phase approximation 
(phRPA)~\cite{bohmCollectiveDescriptionElectron1951,renRandomphaseApproximationIts2012},
ppRPA describes the response of the pairing matrix to a perturbation in the form of a pairing field,
which conveys information in the particle-particle and the hole-hole channel.
For ground states,
the ppRPA correlation energy is shown to be equivalent to 
the ladder-coupled-cluster doubles\cite{pengEquivalenceParticleparticleRandom2013,scuseriaParticleparticleQuasiparticleRandom2013}.
In addition,
ppRPA is the first known functional that captures the energy derivative discontinuity 
in strongly correlated systems\cite{vanaggelenExchangecorrelationEnergyPairing2013}.
For excited-state properties,
the two-electron addition energy and the two-electron removal energy are obtained 
by solving the ppRPA working equation instead of the excitation energy in the particle-hole channel.
Therefore,
the excitation energy of the $N$-electron system can be obtained in two different manners: 
a) the particle-particle channel with the difference between two-electron addition energies of the ($N-2$)-electron system\cite{yangDoubleRydbergCharge2013},
and 
b) the hole-hole channel with the difference between two-electron addition energies of the ($N+2$)-electron system\cite{yangDoubleRydbergCharge2013,bannwarthHoleHoleTamm2020,yuInitioNonadiabaticMolecular2020},
where $N$ is the number of electrons.
It has been shown that ppRPA predicts accurate excited-state properties of molecular systems, 
including valence, double, Rydberg and charge transfer  excitation energies\cite{yangBenchmarkTestsSpin2013,
yangDoubleRydbergCharge2013,
yangExcitationEnergiesParticleparticle2014,
yangChargeTransferExcitations2017,al-saadonAccurateTreatmentChargeTransfer2018,
zhangAccurateEfficientCalculation2016,
liLinearScalingCalculations2023},
analytic gradients\cite{zhangAnalyticGradientsGeometry2014},
conical intersections\cite{yangDoubleRydbergCharge2013}, 
oscillator strengths\cite{yangConicalIntersectionsParticle2016},
and spin-state energetics\cite{pinterSpinstateEnergeticsIron2018}.
Recently, ppRPA with the Tamm-Dancoff approximation (TDA) has been applied 
in the multi-reference DFT approach to describe dissociation breakings and to predict excitation energies\cite{
chenMultireferenceDensityFunctional2017,
liMultireferenceDensityFunctional2022}.
ppRPA is also used in the Green's function formalism,
where the self-energy in the T-matrix approximation is formulated with the ppRPA eigenvalues 
and eigenvectors to calculate quasiparticle energies\cite{
zhangAccurateQuasiparticleSpectra2017,
liRenormalizedSinglesGreen2021,
liLinearScalingCalculations2023,
orlandoThreeChannelsManybody2023,
loosStaticDynamicBethe2022}.

The computational cost of the ppRPA approach can also be significantly reduced using 
the active-space formalism~\cite{liLinearScalingCalculations2023,zhangAccurateEfficientCalculation2016}.
In the recently developed active-space formalism,
only the particle and hole pairs with large contributions to low-lying excitation energies 
are included by constraining both indices in particle and hole pairs.
It has been shown that using an active space of only $30$ occupied and $30$ virtual orbitals, 
active-space ppRPA achieves fast convergence to within $0.05$ \,{eV} compared to full ppRPA 
for molecular excitations of different characters, 
including charge-transfer, Rydberg, double, and valence excitations as well as diradicals\cite{liLinearScalingCalculations2023}.  As a result, the ppRPA calculations
for molecular excitations becomes linear scaling and is more efficient than the ground state SCF calculations of the same molecules.

In this work, 
we apply the active-space ppRPA approach to predict vertical excitation energies in 
solid-state point defects,
including the negatively charged nitrogen-vacancy center (NV$^-$) 
and the neutral silicon-vacancy center (SiV$^0$) in diamond, 
and the $kk$-configuration of the neutral divacancy center ($kk$-VV$^0$) in 4H silicon carbide (4H-SiC).
Here, the ($N+2$)-electron ground state is computed with DFT and is used as the reference in ppRPA calculations.
By adding two electrons to the original $N$-electron defect system,
the ground state becomes closed-shell,
which can be straightforwardly described by single-determinant Kohn-Sham DFT.
All desired excitation energies can then be obtained by taking the differences 
between two-electron removal energies of the ($N+2$)-electron system.
We demonstrate that the excitation energy converges rapidly with respect to the size of the active space 
using supercells of various sizes.
With a small active space consisting of only 200 occupied and 200 virtual orbitals, 
ppRPA predicts accurate excitation energies for all tested defect systems. 
To the best of our knowledge, 
this work is the first application of ppRPA for excitation energies in realistic periodic bulk systems.

We first review the ppRPA formalism.
As the counterpart of phRPA formulated in the particle-hole channel,
ppRPA is formulated with the particle-particle propagator that completely describes 
the dynamic fluctuation of the pairing matrix\cite{
vanaggelenExchangecorrelationEnergyPairing2013,
vanaggelenExchangecorrelationEnergyPairing2014}.
In the frequency space, 
the time-ordered pairing matrix fluctuation is\cite{
vanaggelenExchangecorrelationEnergyPairing2013,
vanaggelenExchangecorrelationEnergyPairing2014} 
\begin{equation}\label{eq:pairing_matrix}
    K_{pqrs}(\omega) = 
        \sum_m \frac{
            \langle \Psi^N_0 | \hat{a}_p \hat{a}_q | \Psi^{N+2}_0 \rangle 
            \langle \Psi^{N+2}_0 | \hat{a}_s^{\dagger} \hat{a}_r^{\dagger} | \Psi^N_0 \rangle 
            }{
            \omega - \Omega^{N+2}_m + i\eta
            }
        - \sum_m \frac{
            \langle \Psi^N_0 | \hat{a}_s^{\dagger} \hat{a}_r^{\dagger} | \Psi^{N-2}_0 \rangle 
            \langle \Psi^{N-2}_0 | \hat{a}_p \hat{a}_q | \Psi^N_0 \rangle 
            }{
            \omega - \Omega^{N-2}_m - i\eta 
            }
\end{equation}
where $\hat{a}_p^{\dagger}$ and $\hat{a}_p$ are the second quantization creation and annihilation operators, 
$\Omega^{N\pm2}$ is the two-electron addition/removal energy, 
and $\eta$ is a positive infinitesimal number.
In Eq.~\ref{eq:pairing_matrix} and the following, 
we use $i$, $j$, $k$, $l$ for occupied orbitals, 
$a$, $b$, $c$, $d$ for virtual orbitals, 
$p$, $q$, $r$, $s$ for general molecular orbitals, 
and $m$ for the index of the two-electron addition/removal energy.

Similar to the phRPA,
the pairing matrix fluctuation $K$ of the interacting system can be approximated 
from the non-interacting $K_0$ with the Dyson equation\cite{
vanaggelenExchangecorrelationEnergyPairing2013,
vanaggelenExchangecorrelationEnergyPairing2014}
\begin{equation}\label{eq:dyson}
    K = K^0 + K^0 V K
\end{equation}
where the antisymmetrized interaction 
$V_{pqrs}
    = \langle pq || rs \rangle 
    = \langle pq | rs \rangle - \langle pq | sr \rangle$ 
is used and 
$\langle pq|rs\rangle 
    = \int dx_{1}dx_{2}
        \frac{
        \phi_{p}^{*}(x_{1})\phi_{q}^{*}(x_{2})\phi_{r}(x_{1})\phi_{s}(x_{2})
        }{
        |r_{1}-r_{2}|
        }$.
The direct ppRPA can be obtained by neglecting the exchange term in $V$ in Eq.~\ref{eq:dyson}\cite{tahirComparingParticleparticleParticlehole2019}. 

Eq.~\ref{eq:dyson} can be cast into a generalized eigenvalue equation,
which is similar to the Casida equation in TDDFT\cite{casidaTimeDependentDensityFunctional1995,ullrichTimeDependentDensityFunctionalTheory2011}
\begin{equation}\label{eq:eigen_equation}
\begin{bmatrix}\mathbf{A} & \mathbf{B}\\
\mathbf{B}^{\text{T}} & \mathbf{C}
\end{bmatrix}\begin{bmatrix}\mathbf{X}\\
\mathbf{Y}
\end{bmatrix}=\Omega^{N\pm2}\begin{bmatrix}\mathbf{I} & \mathbf{0}\\
\mathbf{0} & \mathbf{-I}
\end{bmatrix}\begin{bmatrix}\mathbf{X}\\
\mathbf{Y}
\end{bmatrix}
\end{equation}
with
\begin{align}
A_{ab,cd} & =\delta_{ac}\delta_{bd}(\epsilon_{a}+\epsilon_{b})+\langle ab||cd\rangle \\
B_{ab,kl} & =\langle ab||kl\rangle \\
C_{ij,kl} & =-\delta_{ik}\delta_{jl}(\epsilon_{i}+\epsilon_{j})+\langle ij||kl\rangle 
\end{align}
where $a<b$, $c<d$, $i<j$, $k<l$ and $\Omega^{N\pm2}$ is the two-electron addition/removal energy.
In our ppRPA calculations of defect systems,
the DFT self-consistent field (SCF) calculation of the ($N+2$)-electron state at 
the ground-state geometry of $N$-electron system is first performed,
then the orbital energies and orbitals are used in the working equation Eq.~\ref{eq:eigen_equation} for calculating two-electron removal energies.
The excitation energy can be obtained from the difference between the lowest and a higher two-electron removal energy.

Since the corresponding ($N+2$)-electron system is closed-shell,
Eq.~\ref{eq:eigen_equation} can be expressed in the spin-adapted form\cite{yangBenchmarkTestsSpin2013}.
The singlet ppRPA matrix is given by 
\begin{align}
A^{\text{s}}_{ab,cd} & =\delta_{ac}\delta_{bd}(\epsilon_{a}+\epsilon_{b})+\langle ab||cd\rangle \label{eq:a_singlet}\\
B^{\text{s}}_{ab,kl} & =\langle ab||kl\rangle \label{eq:b_singlet}\\
C^{\text{s}}_{ij,kl} & =-\delta_{ik}\delta_{jl}(\epsilon_{i}+\epsilon_{j})+\langle ij||kl\rangle \label{eq:c_singlet} 
\end{align}
with  $a<b$, $c<d$, $i<j$ and $k<l$.
And the triplet ppRPA matrix is given by
\begin{align}
A^{\text{t}}_{ab,cd} & =\delta_{ac}\delta_{bd}(\epsilon_{a}+\epsilon_{b}) 
+ \frac{1}{\sqrt{(1+\delta_{ab})(1+\delta_{cd})}} (\langle ab|cd\rangle + \langle ab|dc\rangle) \label{eq:a_triplet}\\
B^{\text{t}}_{ab,kl} & = \frac{1}{\sqrt{(1+\delta_{ab})(1+\delta_{kl})}} 
(\langle ab|kl\rangle + \langle ab|lk\rangle) \label{eq:b_triplet}\\
C^{\text{t}}_{ij,kl} & =-\delta_{ik}\delta_{jl}(\epsilon_{i}+\epsilon_{j})
+ \frac{1}{\sqrt{(1+\delta_{ij})(1+\delta_{kl})}} (\langle ij|kl\rangle + \langle ij|lk\rangle) \label{eq:c_triplet}
\end{align}
with  $a \leq b$, $c \leq d$, $i \leq j$ and $k \leq l$.

As introduced in Ref.\citenum{liLinearScalingCalculations2023},
the ppRPA matrix can be constructed in an active space that constrains indices of both particle and hole pairs.
This means that for singlet ppRPA matrix in Eq.~\ref{eq:a_singlet} to Eq.~\ref{eq:c_singlet}, the indices are constrained as
\begin{align}
    & a < b \leq N_{\text{vir,act}} \text{ and } c < d \leq N_{\text{vir,act}}  \\
    & i < j \leq N_{\text{occ,act}} \text{ and } k < l \leq N_{\text{occ,act}}
\end{align}
where $N_\mathrm{occ,act}$ and $N_\mathrm{vir,act}$ are the numbers of occupied and virtual orbitals in the active space. For the triplet ppRPA matrix in Eq.~\ref{eq:a_triplet} to Eq.~\ref{eq:c_triplet}, the indices are constrained in the same way.
As shown in Ref.\citenum{liLinearScalingCalculations2023},
the above active space only includes the particle and hole pairs with large contributions to low-lying excitation energies,
which greatly reduces the computational cost of the ppRPA calculations. The scaling of active-space ppRPA is $\mathcal{O} (N_\mathrm{act}^4)$ using the Davidson algorithm\cite{yangExcitationEnergiesParticleparticle2014,stratmannEfficientImplementationTimedependent1998} with $N_\mathrm{act}$ being a small number of active-space orbitals. 
In this work, 
the full diagonalization was used for ppRPA calculations due to the small size of the active space.
In addition, the AO-to-MO transformation step scales as $\mathcal{O} (N_\mathrm{aux} N^2_\mathrm{AO} N_\mathrm{act})$ with density fitting, where $N_\mathrm{AO}$ and $N_\mathrm{aux}$ are the numbers of computational and auxiliary basis functions.

Ground-state geometries of all three defect systems were optimized with the PBE functional~\cite{perdewGeneralizedGradientApproximation1996} using the Quantum ESPRESSO package~\cite{carnimeoQuantumESPRESSOOne2023,giannozziQuantumESPRESSOExascale2020}, and details can be found in the Supporting Information (SI). We then performed all ($N+2$)-electron ground-state DFT calculations in periodic Gaussian basis sets with Gaussian density fitting using the PySCF quantum chemistry software package~\cite{sunPySCFPythonbasedSimulations2018,sunRecentDevelopmentsPySCF2020}, in supercell calculations with $\Gamma$-point sampling. Two functionals (PBE\cite{perdewGeneralizedGradientApproximation1996} and B3LYP~\cite{beckeDensityFunctionalThermochemistry1993,leeDevelopmentColleSalvettiCorrelationenergy1988}) were used, in combination with the cc-pVDZ basis set~\cite{dunningGaussianBasisSets1989} and the cc-pVDZ-RI auxiliary basis set~\cite{weigendEfficientUseCorrelation2002}. With the electron integrals and DFT results obtained from PySCF, we further performed active-space ppRPA calculations with periodic boundary conditions to predict vertical excitation energies of point defects using the Lib\_ppRPA library\cite{liUnpublished}. Data from ppRPA basis set convergence tests can be found in the SI.

\begin{figure}
\includegraphics[width=0.75\textwidth]{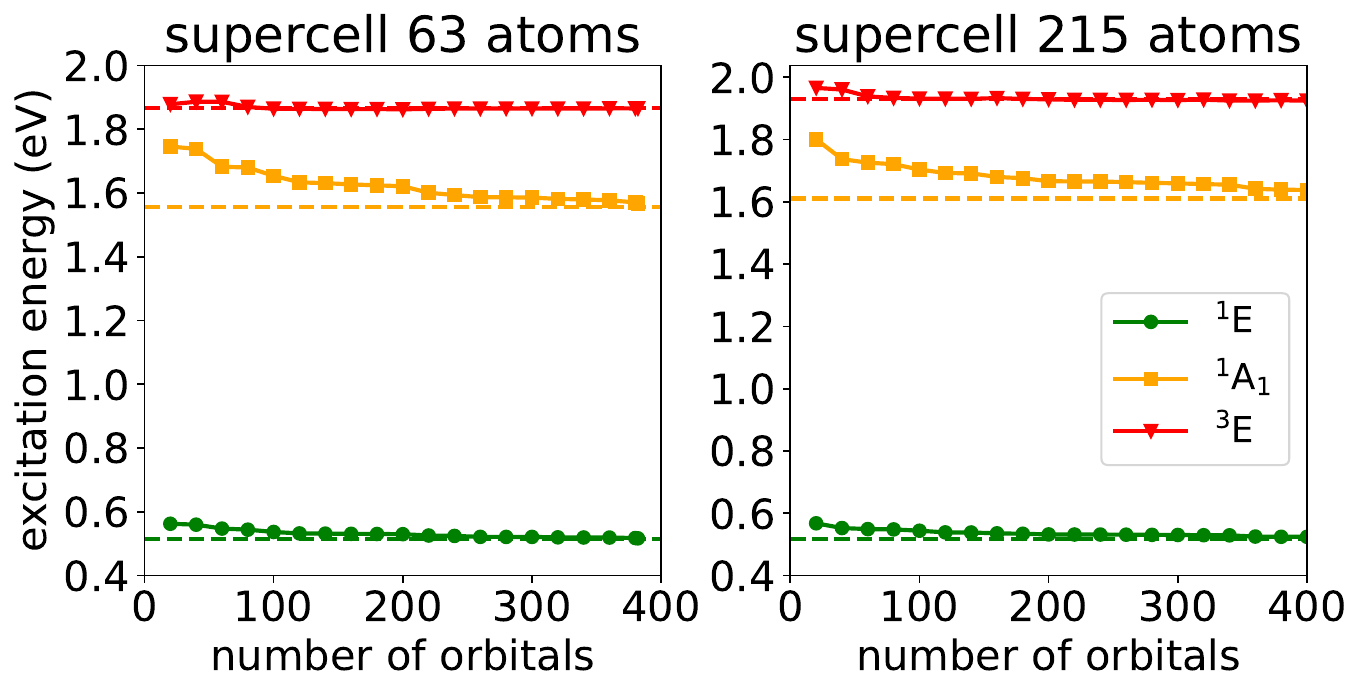}
\caption{Vertical excitation energies of NV$^-$ in diamond obtained from ppRPA@PBE with respect to the number of orbitals in the active space.
The numbers of occupied and virtual orbitals in the active space are the same.
Left: supercell containing $63$ atoms.
Full ($N+2$)-electron state has $191$ occupied and $693$ virtual orbitals.
In the last point,
$191$ occupied and $200$ virtual orbitals are included in the active space.
The dashed lines correspond to excitation energies obtained from the active space with $191$ occupied and $300$ virtual orbitals.
Right: supercell containing $215$ atoms.
Full ($N+2$)-electron state has $647$ occupied and $2363$ virtual orbitals.
The dashed lines correspond to excitation energies obtained from the active space with $300$ occupied and $300$ virtual orbitals.}
\label{fig:convergence}
\end{figure}

We first examine the convergence behavior of the excitation energies obtained from the ppRPA approach with respect to the size of the active space.
The active-space ppRPA@PBE excitation energies of NV$^-$ center using supercells containing $63$ and $215$ atoms are shown in Fig.~\ref{fig:convergence}.
The supercell containing $63$ atoms has $191$ occupied and $691$ virtual orbitals,
while the 215-atom supercell has $647$ occupied and $2363$ virtual orbitals.
For simplicity,
the active spaces used in this work include the same numbers of occupied and virtual orbitals.
As seen in Fig.~\ref{fig:convergence},
the excitation energies of all three excited states (i.e., $^1$E, $^1$A$_1$, and $^3$E) converge rapidly with respect to the size of the active space for both supercell sizes.
For the low-lying singlet excited state $^1$E and triplet excited state $^3$E,
using an active space consisting of only $30$ occupied and $30$ virtual orbitals gives errors of only $0.03$ \,{eV} compared with much larger active-space results (i.e., 191 occupied and 300 virtual orbitals),
which is the same as the convergence behavior for excitation energies in molecular systems reported in Ref.\citenum{liLinearScalingCalculations2023}.
For the $^1$A$_1$ excited state, slightly slower convergence is observed. Nevertheless, the ppRPA@PBE $^1$A$_1$ excitation energy is converged to within $0.03$ \,{eV} when using $200$ (191 for 63-atom supercell) occupied and $200$ virtual orbitals. Similar behaviors have been observed for ppRPA@B3LYP calculations and for other defect systems, as shown in the SI. Thus, we choose to use the active space composed of $200$ occupied and $200$ virtual orbitals in all ppRPA calculations in this work.

\renewcommand{\arraystretch}{1.2}
\begin{table}[h!]
\setlength\tabcolsep{15pt}
\caption{Vertical excitation energies of NV$^-$ in diamond obtained from the ppRPA approach based on PBE and B3LYP functionals compared with reference values.
Extrapolated values were obtained from the results of supercells containing $63$ and $215$ atoms.
In ppRPA calculations, the geometry of the $^3$A$_2$ ground state and cc-pVDZ basis set were used. 
All values are in \,{eV}.
}\label{tab:nv}
\begin{tabular}{c|ccc}
\hline\hline
                            & $^1$E          & $^1$A$_1$       & $^3$E \\
\hline
Experiment~\cite{daviesOpticalStudies9451997,haldarLocalExcitationsCharged2023}                        & $0.50\sim0.59$ & $1.76\sim1.85$  & 2.18   \\
ppRPA@PBE (supercell 215)                              & 0.52           & 1.64            & 1.92   \\
ppRPA@PBE (extrapolated)                               & 0.53           & 1.67            & 1.95   \\
ppRPA@B3LYP (supercell 215)                            & 0.60           & 1.92            & 2.09   \\
ppRPA@B3LYP (extrapolated)                             & 0.61           & 1.97            & 2.10   \\\hline
TDDFT@PBE~\cite{jinExcitedStateProperties2023}         & 0.51           & 1.34            & 2.09   \\
TDDFT@DDH~\cite{jinExcitedStateProperties2023}         & 0.68           & 1.97            & 2.37   \\
$G_0W_0$-BSE@PBE~\cite{maExcitedStatesNegatively2010}  & 0.40           & 0.99            & 2.32   \\
DMET-NEVPT2~\cite{haldarLocalExcitationsCharged2023}   & 0.53           & 1.62            & 2.40   \\
QDET~\cite{jinExcitedStateProperties2023}              & 0.48           & 1.32            & 2.16   \\
CRPA+CI~\cite{bockstedteInitioDescriptionHighly2018}   & 0.49           & 1.41            & 2.02   \\     
\hline\hline
\end{tabular}
\end{table}

The vertical excitation energies (VEEs) of NV$^-$ in diamond obtained from the ppRPA approach based on PBE and B3LYP using the 215-atom supercell and a two-point extrapolation scheme are presented in Table~\ref{tab:nv}. The extrapolation was done using 63-atom and 215-atom excitation energies in a linear fitting of the form: $E(1/N_\mathrm{atom}) = E_\infty + a/N_\mathrm{atom}$.
We note that, 
there are uncertainties in the estimation of experimental VEEs. 
For example, 
when based on experimental zero-phonon lines (ZPLs) and the Franck-Condon shifts obtained from TDDFT calculations~\cite{jinExcitedStateProperties2023}, 
the VEEs are reported to be $0.40\sim 0.55$ \,{eV} for the $^1$E state and $1.53\sim 1.62$ \,{eV} for the $^1$A$_1$ state.
It is well-known that TDDFT has an undesired dependence on the exchange-correlation functional. In Table~\ref{tab:nv}, TDDFT@PBE severely underestimates the excitation energy of the $^1$A$_1$ state, while TDDFT@DDH overestimates excitation energies of all excited states by $0.1 \sim 0.2$ \,{eV}.
$G_0W_0$-BSE@PBE underestimates the excitation energy of the $^1$A$_1$ state by $0.8$ \,{eV} and has errors around $0.1$ \,{eV} for other two states.
The embedding approaches, including DMET, QDET, and CRPA+CI, predict accurate excitation energies for low-lying $^1$E and $^3$E states.
However, QDET and CRPA+CI significantly underestimate the excitation energy of the $^1$A$_1$ state by $0.4$ \,{eV} or more.
In contrast, we find that the ppRPA approach provides a balanced description of all excited states. For three excited states,
ppRPA@B3LYP predicts accurate excitation energies with errors around $0.1$ \,{eV}.
The ppRPA approach based on PBE slightly underestimates excitation energies, especially for $^1$A$_1$ and $^3$E states, with errors around $0.2$ \,{eV}.
In addition,
ppRPA has a weaker DFT starting-point dependence than TDDFT.
For instance,
the difference between excitation energies of the $^1$A$_1$ state obtained from ppRPA with the GGA and the hybrid functionals is only $0.3$ \,{eV},
which is half of the $0.6$ \,{eV} difference between TDDFT@DDH and TDDFT@PBE.
As shown in Table~\ref{tab:nv} and in the SI,
we also observe that excitation energies for NV$^-$ in diamond have a weak dependence on the size of the supercell model.
The difference between ppRPA excitation energies from using the 215-atom supercell and those from the two-point extrapolation is smaller than $0.05$ \,{eV}.

\begin{table}[h!]
\setlength\tabcolsep{20pt}
\caption{Vertical excitation energies of SiV$^0$ in diamond obtained from the ppRPA approach based on PBE and B3LYP functionals compared with reference values. 
Experimental vertical excitation energy is estimated by combining the experimental ZPL value of 1.31 eV~\cite{daviesOpticalStudies9451997} and the Franck-Condon shift of 0.29 eV from the TDDFT calculation~\cite{jinExcitedStateProperties2023}.
Extrapolated values were obtained from the results of supercells containing 63 and 215 atoms.
In ppRPA calculations, 
the geometry of the $^3$A$_{2g}$ ground state and cc-pVDZ basis set were used. 
All values are in \,{eV}.
}\label{tab:siv}
\begin{tabular}{c|ccc}
\hline\hline
                          & $^3$A$_{2u}$ & $^3$E$_u$  & $^3$A$_{1u}$ \\
\hline
Experiment                                           &              & 1.60       &              \\
ppRPA@PBE (supercell 215)                            & 1.47         & 1.54       & 1.81         \\
ppRPA@PBE (extrapolated)                             & 1.27         & 1.30       & 1.54         \\
ppRPA@B3LYP (supercell 215)                          & 1.54         & 1.62       & 1.89         \\
ppRPA@B3LYP (extrapolated)                           & 1.34         & 1.38       & 1.62         \\ \hline
TDDFT@PBE~\cite{jinExcitedStateProperties2023}       & 1.24         & 1.28       & 1.37         \\
TDDFT@DDH~\cite{jinExcitedStateProperties2023}       & 1.49         & 1.57       & 1.76         \\
DMET-CASSCF~\cite{mitraExcitedStatesCrystalline2021} & 2.26         & 2.44       & 3.16         \\
DMET-NEVPT2~\cite{mitraExcitedStatesCrystalline2021} & 2.39         & 2.47       & 2.61         \\
QDET~\cite{maQuantumEmbeddingTheory2021}             &              & 1.59       & 1.62         \\
\hline\hline
\end{tabular}
\end{table}

We now turn to the discussion of SiV$^0$ in diamond.
The vertical excitation energies of SiV$^0$ in diamond obtained from the ppRPA approach based on PBE and B3LYP using the 215-atom supercell and those from the two-point extrapolation are presented in Table~\ref{tab:siv}.
Here, TDDFT@DDH achieves high accuracy for predicting the excitation energy of the $^3$E$_u$ state. On the other hand, TDDFT@PBE underestimates the $^3$E$_u$ energy by 0.3 eV.
Similar to the prediction of NV$^-$ in diamond,
TDDFT shows a starting-point dependence for SiV$^0$ in diamond, where
the difference in excitation energies from different functionals can be as large as $0.4$ \,{eV}.
Among quantum embedding approaches,
the QDET approach produces the smallest error of only $0.02$ \,{eV}.
The large errors in DMET-CASSCF and DMET-NEVPT2 may be attributed to the finite-size error and the unsatisfactory treatment of the hybridization between defect orbitals and the environment\cite{mitraExcitedStatesCrystalline2021}.
With the supercell model containing $215$ atoms,
both ppRPA@PBE and ppRPA@B3LYP predict accurate excitation energies for the $^3$E$_u$ state with errors smaller than $0.05$ \,{eV}.
However, the extrapolated results give slightly larger errors around $0.2 \sim 0.3$ \,{eV}.
Compared with TDDFT,
the starting-point dependence in ppRPA is largely reduced.
The differences of excitation energies obtained from ppRPA based on PBE and B3LYP are smaller than $0.1$ \,{eV}.
Similar to NV$^-$ in diamond,
ppRPA using the hybrid functional B3LYP provides better accuracy than ppRPA based on the GGA functional PBE for SiV$^0$ in diamond.

\begin{table}[h]
\setlength\tabcolsep{25pt}
\caption{Vertical excitation energies of $kk$-VV$^0$ in 4H-SiC obtained from the ppRPA approach based on PBE and B3LYP using the 286-atom supercell.
Experimental vertical excitation energy is estimated by combining the experimental ZPL value of 1.10 eV~\cite{jinPhotoluminescenceSpectraPoint2021} and the Franck-Condon shift of 0.11 eV from the TDDFT calculation~\cite{jinExcitedStateProperties2023}.
In ppRPA calculations, the geometry of the $^3$A$_2$ ground state and cc-pVDZ basis set were used. All values are in \,{eV}.
}\label{tab:vv}
\begin{tabular}{c|ccc}
\hline\hline
                          & $^1$E & $^1$A$_1$ & $^3$E \\
\hline
Experiment                                           &       &           & 1.21  \\
ppRPA@PBE (supercell 286)                            & 0.28  & 0.88      & 1.12  \\
ppRPA@B3LYP (supercell 286)                          & 0.33  & 1.13      & 1.23  \\ \hline
TDDFT@PBE~\cite{jinExcitedStateProperties2023}       & 0.33  & 0.90      & 1.41  \\
TDDFT@DDH~\cite{jinExcitedStateProperties2023}       & 0.42  & 1.41      & 1.46  \\
CRPA+CI~\cite{bockstedteInitioDescriptionHighly2018} & 0.29  & 0.88      & 1.13  \\
\hline\hline
\end{tabular}
\end{table}

We also applied the ppRPA approach to calculate excitation energies of $kk$-VV$^0$ in 4H-SiC.
The vertical excitation energies of the $kk$-configuration of VV$^0$ in 4H-SiC obtained from the ppRPA approach based on PBE and B3LYP are shown in Table~\ref{tab:vv}.
TDDFT with PBE or DDH functional overestimates the excitation energy of the $^3$E state by around $0.2$ \,{eV}.
Compared with TDDFT,
CRPA+CI provides improved accuracy with an underestimation of 0.1 eV for the excitation energy of the $^3$E state.
As shown in Ref.\citenum{davidssonFirstPrinciplesPredictions2018}, it is more challenging to extrapolate the excitation energies of $kk$-VV$^0$ in 4H-SiC with respect to the supercell size.
Thus, we only include 
excitation energies obtained from ppRPA using the supercell model containing $286$ atoms in Table~\ref{tab:vv}.
ppRPA@PBE shows similar results to CRPA+CI with an error of $0.1$ \,{eV}.
ppRPA@B3LYP gives the best prediction for the $^3$E state with an error of only $0.02$ \,{eV}. We note that, however, with finite size extrapolation, the ppRPA-predicted $^3$E excitation energy will likely decrease slightly.

\begin{figure}[hbt]
\includegraphics[width=0.8\textwidth]{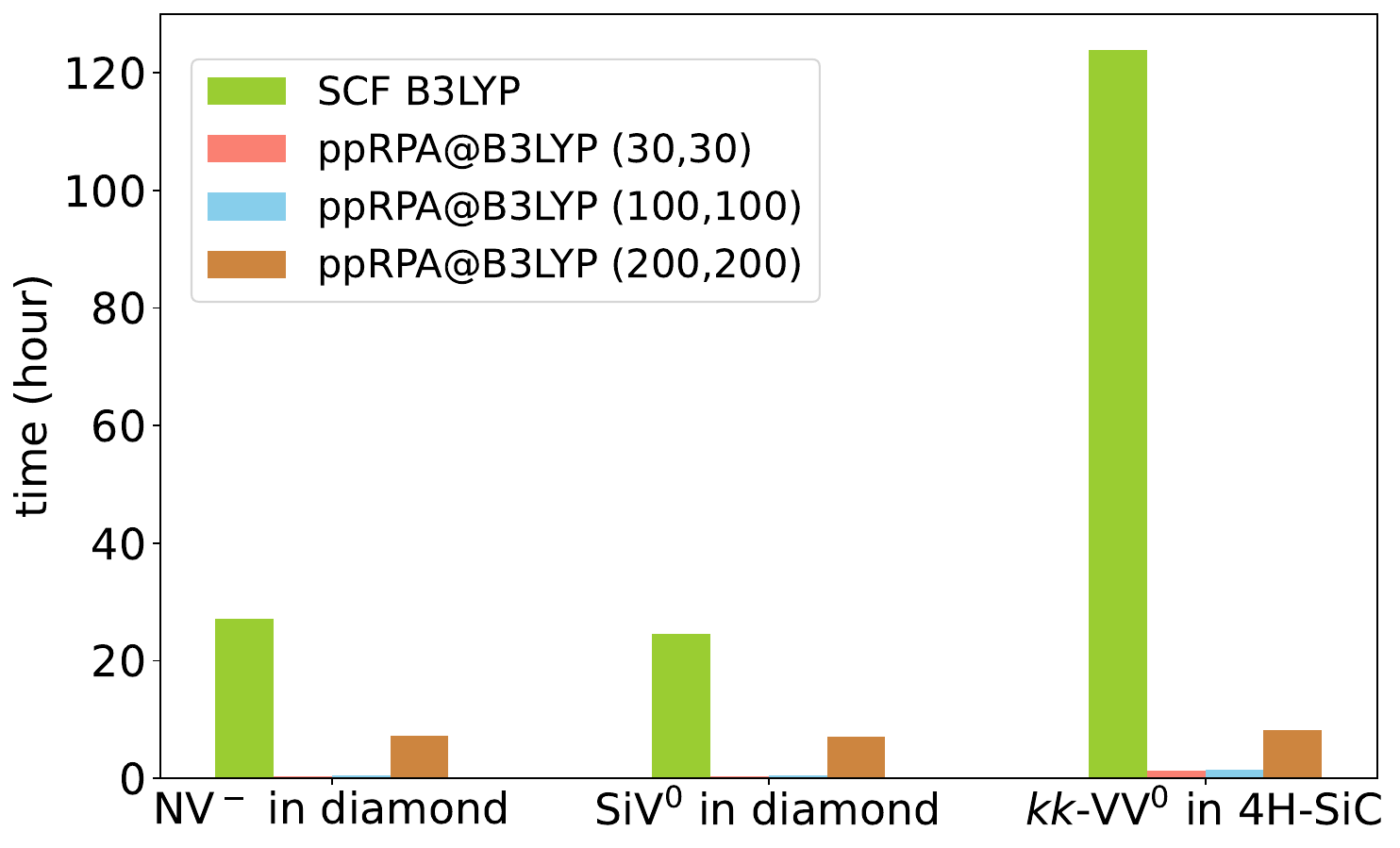}
\caption{Wall time comparison for the ground-state B3LYP SCF calculations and ppRPA@B3LYP calculations with different active spaces. All calculations were performed on a 48-core CPU node. The supercell model containing $215$, $215$, and $286$ atoms was used for NV$^-$ in diamond, SiV$^0$ in diamond, and $kk$-VV$^0$ in 4H-SiC, respectively. ($O$, $V$) means that $O$ occupied orbitals and $V$ virtual orbitals were used in the active space.}
\label{fig:time}
\end{figure}

In addition to good accuracy for predicting excitation energies of point defects,
ppRPA combined with the active-space formalism has a favorable computational scaling.
The wall time comparison for the ground-state B3LYP SCF calculations and ppRPA@B3LYP calculations with different active spaces are demonstrated in Fig.~\ref{fig:time}.
The computational cost for ppRPA with $60$ or $200$ orbitals (including AO to MO transformation and ppRPA steps) is negligible compared to the ground-state SCF for the ($N+2$)-electron system. At such low cost, reasonably accurate excitation energy prediction can already be obtained from the ppRPA calculations (e.g., with 100 occupied and 100 virtual orbitals).
Even for a larger active space with $200$ occupied and $200$ virtual orbitals,
the computational cost of ppRPA is still much smaller than the ground-state DFT calculation.
It is also seen that the computational cost of converged active-space ppRPA calculation is nearly independent of the supercell size.
For the tested defect systems, the ($N+2$)-electron systems become closed-shell by adding two electrons to half-occupied defect orbitals, which only need to be described by a spin-restricted calculation. This means that the DFT calculation here is cheaper than spin-unrestricted DFT/HF typically used in TDDFT and quantum embedding approaches, although we do not claim such acceleration is universal in all point defect systems. We would like to point out that ppRPA calculations of $(N+2)$- or $(N-2)$-excited states can be reviewed as an seamless Fock space embedding approach of capturing explicitly the correlated interactions of two particles or two holes in the medium of the $N$-electron system described with a density functional approximation\cite{zhangAccurateEfficientCalculation2016}. 
It is also interesting to point out that while screening is critical for electron interaction in bulk systems, ppRPA, without the pairing interaction kernel, ignores screening\cite{pengLinearresponseTimedependentDensityfunctional2014}.
Yet ppRPA is shown here to describe the excitation of point defect well. This is likely due to the cancellation of error, as the ground and excited states of the two particle states have similar localization and screening.

In summary, 
we applied the ppRPA approach to predict accurate excitation energies of point defects in semiconductors and insulators. 
In ppRPA calculations,
the ground-state SCF calculation for ($N+2$)-electron system is first performed,
then excitation energies can be efficiently obtained with an active space consisting only $200$ occupied and $200$ virtual orbitals. 
We demonstrated that ppRPA provides a balanced description for different excited states in all tested defect systems including NV$^-$ in diamond, SiV$^0$ in diamond and $kk$-VV$^0$ in 4H-SiC.
The errors from ppRPA@B3LYP for predicting excitation energies of the tested point defects are consistently smaller than $0.2$ \,{eV}.
Furthermore, the computational cost of ppRPA is negligible compared with the ground-state DFT calculation when using the active-space formalism.
Therefore, we conclude that ppRPA shows promise as an accurate and low-cost tool for investigating excited-state properties of point defect systems.
This work also opens up new opportunities for the application of the ppRPA approach to periodic systems.

\begin{acknowledgement}
T.Z. and J.L. are supported by the National Science Foundation (Grant No.~CHE-2337991) and a start-up fund from Yale University. J.L. also acknowledges support from the Tony Massini Postdoctoral Fellowship in Data Science. Y.J. acknowledges support from the Midwest Integrated Center for Computational Materials (MICCoM) as part of the Computational Materials Science Program funded by the U.S. Department of Energy, Office of Science, Office of Basic Energy Sciences, under Contract No.~DE-AC02-06CH11357. J.Y. and W.Y. acknowledge the support from the National Science Foundation (Grant No.~CHE-2154831).
\end{acknowledgement}

\begin{suppinfo}
Details about geometry optimization,
numerical results of excitation energies obtained from ppRPA in defect systems,
basis set convergence for excitation energies obtained from ppRPA.
\end{suppinfo}

\bibliography{ref,unpublished}

\end{document}